\documentclass{ecai} 
\usepackage{latexsym}
\usepackage{amssymb}
\usepackage{amsmath}
\usepackage{amsthm}
\usepackage{booktabs}
\usepackage{enumitem}
\usepackage{graphicx}
\usepackage{color}
\usepackage{multicol}
\usepackage{multirow}
\usepackage{url}

\newcommand{\BibTeX}{B\kern-.05em{\sc i\kern-.025em b}\kern-.08em\TeX}

\begin{document}

%%%%%%%%%%%%%%%%%%%%%%%%%%%%%%%%%%%%%%%%%%%%%%%%%%%%%%%%%%%%%%%%%%%%%%%%

\begin{frontmatter}

%%% Use this command to specify your submission number.
%%% In doubleblind mode, it will be printed on the first page.

\paperid{5640} 

%%% Use this command to specify the title of your paper.

\title{TRUCE-AV: A Multimodal dataset for Trust and Comfort Estimation in Autonomous Vehicles}

%%% Use this combinations of commands to specify all authors of your 
%%% paper. Use \fnms{} and \snm{} to indicate everyone's first names 
%%% and surname. This will help the publisher with indexing the 
%%% proceedings. Please use a reasonable approximation in case your 
%%% name does not neatly split into "first names" and "surname".
%%% Specifying your ORCID digital identifier is optional. 
%%% Use the \thanks{} command to indicate one or more corresponding 
%%% authors and their email address(es). If so desired, you can specify
%%% author contributions using the \footnote{} command.

\author[A,B]{\fnms{Aditi}~\snm{Bhalla}}
\author[B]{\fnms{Christian}~\snm{Hellert}}
\author[A]{\fnms{Enkelejda}~\snm{Kasneci}}
\author[B]{\fnms{Nastassja}~\snm{Becker}}

\address[A]{Technical University of Munich, Germany}
\address[B]{Continental Automotive Technologies GmbH, Germany}

%%% Use this environment to include an abstract of your paper.

\begin{abstract}

Understanding and estimating driver trust and comfort are essential for the safety and widespread acceptance of autonomous vehicles. Existing works analyze user trust and comfort separately, with limited real-time assessment and insufficient multimodal data. This paper introduces a novel multimodal dataset called TRUCE-AV, focusing on trust and comfort estimation in autonomous vehicles. The dataset collects real-time trust votes and continuous comfort ratings of 31 participants during a simulator-based fully autonomous driving. Simultaneously, physiological signals, such as heart rate, gaze, and emotions, along with environmental data (e.g., vehicle speed, nearby vehicle positions, and velocity), are recorded throughout the drives. Standard pre- and post-drive questionnaires were also administered to assess participants' trust in automation and overall well-being, enabling the correlation of subjective assessments with real-time responses. To demonstrate the utility of our dataset, we evaluated various machine learning models for trust and comfort estimation using physiological data. Our analysis showed that tree-based models like Random Forest and XGBoost and non-linear models such as KNN and MLP regressor achieved the best performance for trust classification and comfort regression. Additionally, we identified key features that contribute to these estimations by using SHAP analysis on the top-performing models. Our dataset enables the development of adaptive AV systems capable of dynamically responding to user trust and comfort levels non-invasively, ultimately enhancing safety, user experience, and human-centered vehicle design. 
\end{abstract}

\end{frontmatter}

%%%%%%%%%%%%%%%%%%%%%%%%%%%%%%%%%%%%%%%%%%%%%%%%%%%%%%%%%%%%%%%%%%%%%%%%

\section{Introduction}

Autonomous vehicles (AVs) are expected to revolutionize mobility systems by reducing accidents, minimizing driving workload, and enhancing travel comfort~\cite{wang2020safety}. However, for these benefits to materialize, people must trust and accept these technologies~\cite{hartwich2021improving}. According to the J.D. Power 2024 U.S. Mobility Confidence Index (MCI) study~\cite{JDPower2024study}, customers show low confidence in AVs, with comfort being one of the most influential factors. Therefore, many studies have been conducted to understand the influence of driver trust and comfort on the widespread adoption of AVs~\cite{choi2015investigating,korber2018introduction,siebert2013discomfort}. According to existing literature, \textbf{Trust} influences user's willingness to rely on an automated system~\cite{lee2004trust, korber2018introduction}, while \textbf{Comfort} is associated with a feeling of well-being~\cite{de2003sitting, carsten2019can}. In other words, trust emphasizes on the user's and vehicle's \textit{actions}; for instance, distrust will lead to taking over the control of the vehicle, and excessive trust can result in safety issues, whereas comfort addresses the user's \textit{subjective feelings} in driving a vehicle. 

Researchers have extensively examined the key factors influencing driver trust~\cite{he2022modelling,stapel2022road,zhang2022trust,naiseh2025trust} and comfort in AVs~\cite{paschalidis2020deriving, meng2024study, bellem2018comfort, aledhari2023motion}. Trust is shaped by factors such as system reliability, transparency of decision-making, and user familiarity or prior experience with automation. Psychological factors, such as dispositional trust and perceived control, further contribute to a user's willingness to rely on AVs. On the other hand, comfort is largely affected by the smoothness of the driving experience, predictability of the vehicle's maneuvers, ride quality, and environmental conditions such as noise and vibration. The presumption that trust and acceptance are prerequisites for system use, but comfort may act as a barrier to adoption~\cite{bellem2018comfort}, has led to further studies investigating the relationship between trust and comfort in the context of AVs~\cite{elbanhawi2015passenger}. According to~\cite{kaufman2025improving, peng2023conceptualising}, trust and comfort are positively correlated, demonstrating that trust enhances comfort by reducing perceived risks, and in turn, comfort increases trust by affirming the system's predictability and safety. Collectively, these studies underline the implicit relationship between trust, comfort, perceived risk, safety, and acceptance of AVs. Therefore, evaluating trust and comfort simultaneously offers a more comprehensive understanding of the user's experience with AVs, thereby assisting in the design of user-centered systems. 

Despite extensive research, studies on trust and comfort in AVs are fragmented, usually isolated, limited in modalities, and lack open-source datasets. The absence of a publicly available dataset limits the ability to compare models, validate findings, and reproduce results across studies. Therefore, we introduce a comprehensive, multi-modal TRUCE-AV (TRUst and Comfort Estimation in AV) dataset\footnote{The dataset can be found at: \url{https://truceav.github.io/}} that captures the temporal evolution of trust and comfort in an immersive simulator-based fully autonomous driving setup. Our dataset includes:

\begin{itemize}
    \item Two driving sessions with diverse scenarios that reflect real-world driving events.
    \item Real-time event-specific trust ratings and continuous comfort measurements from 31 participants.
    \item Concurrent physiological data, such as heart rate, gaze patterns, and emotional response, along with environmental data, such as vehicle speed, distance to nearby vehicles, and weather conditions.
\end{itemize}

\begin{figure}[tb]
        \centering
        \includegraphics[width=0.95\linewidth]{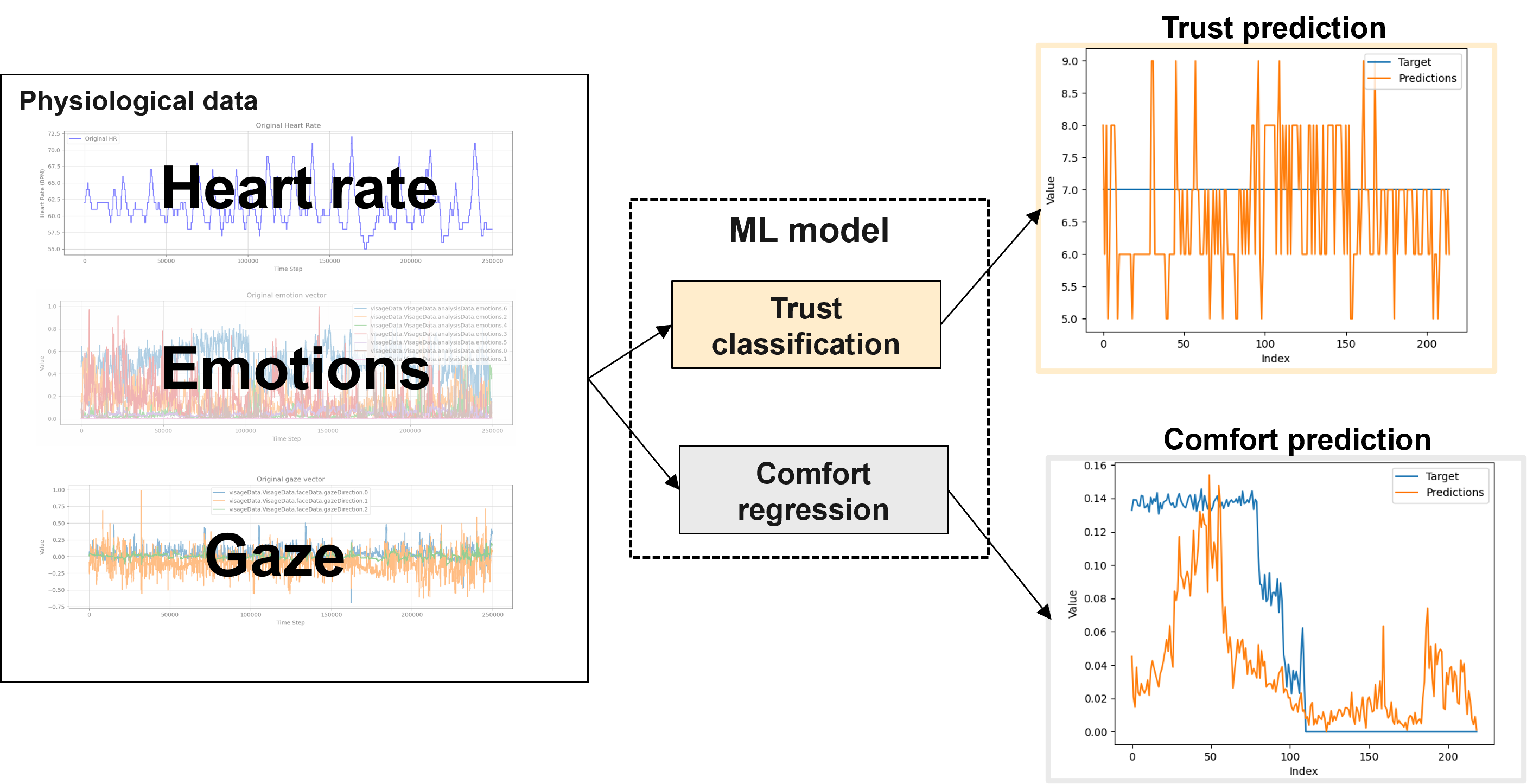}
        \vspace{-8pt}
        \caption{Trust and comfort prediction using physiological features such as heart rate, emotions, and gaze}
        \label{fig:ML model}
   \end{figure}

Additionally, we conducted pre- and post-drive questionnaires to assess each participant's trust, well-being, and comfort. We established baseline results using standard machine learning (ML) models to estimate trust and comfort from physiological data (Fig.\@\ref{fig:ML model}) and identified key contributing features using SHapley Additive exPlanations (SHAP).
%%%%%%%%%%%%%%%%%%%%%%%%%%%%%%%%%%%%%%%%%%%%%%%%%%%%%%%%%%%%%%%%%%%%%%%%

\section{Related Work}

Estimating driver trust and comfort in AVs presents significant challenges due to their complex, subjective, and dynamic nature. Therefore, in this section, we highlight the measurement methods used to collect trust and comfort data and their limitations. We also review the existing datasets and ML-based approaches for trust and comfort estimation.

\paragraph{Trust and comfort measurement:} Several studies measured and evaluated driver trust and comfort in AVs, which can be broadly divided into three categories:  
\begin{itemize}
    \item \textit{Self-reported measures} capture driver's subjective trust~\cite{hunter2022interaction,ayoub2021modeling} and comfort~\cite{norzam2022analysis, su2023estimating} using structured questionnaires or surveys on a Likert-type scale. They are the most commonly used measures to assess people's perception of AVs. However, they cannot capture real-time changes in trust and comfort and are heavily influenced by individual, social, and memory bias. 
    
    \item \textit{Physiological measures} assess trust and comfort indirectly by recording continuous biometric signals such as heart rate (HR), electrocardiography (ECG), galvanic skin response (GSR), electroencephalography (EEG), or eye-tracking, non-invasively. These indicators are a good representation of real-time arousal, stress, or cognitive load that correlates to driver trust~\cite{yi2023human,ayoub2023real} and comfort~\cite{su2021study, beggiato2019physiological, beggiato2020facial}. However, it is challenging to interpret physiological signals due to individual variability.
    
    \item \textit{Behavioral measures} involve observing and systematically recording driver's actions, such as interaction with the AV, takeover performance, and compliance with automation prompts. These task-related interaction behaviors, whether intentional and active or unintentional and passive, are used to infer real-world driver trust~\cite{ayoub2021investigation, cegarra2025driving} and comfort~\cite{paschalidis2020deriving,asarar2021predicting} in AVs. However, these behaviors may be influenced by cognitive overload, prior experience, or uncertainty about the system's limitations.
\end{itemize}
As no single measurement method can accurately capture trust and comfort in AVs, a well-developed study must combine two or more methods. Therefore, in our work, we have used self-reported and physiological measures to incorporate subjective and real-time data for driver trust and comfort estimation in AVs.

\paragraph{Existing datasets}

\begin{table}[ht]
    \centering
    \caption{Publicly available multimodal dataset for trust and comfort estimation in AVs using physiological or environmental features}    \label{tab:dataset}
    \begin{tabular}{p{1.5cm}p{1.5cm}p{1.7cm}p{1cm}}
    \toprule
         \textbf{Ref} &  \multicolumn{2}{c}{\textbf{Modality}} &  \textbf{Ground Truth} \\
         & \textbf{Physiological} & \textbf{Environmental} & \\
    \midrule
    \cite{chen2025predicting}& $\times$ & $\checkmark$ & Comfort  \\
    \cite{meteier2023dataset} & $\checkmark$ & $\times$ & Trust \\
    \cite{comfortdataset} & $\times$ & $\checkmark$ & Comfort  \\
TRUCE-AV & $\checkmark$ & $\checkmark$ &  Trust \& Comfort  \\
    \bottomrule
    \end{tabular}
\end{table}

Despite numerous research studies on driver trust and comfort in AVs, very few have made their datasets publicly available as presented in Table \@\ref{tab:dataset}. The dataset introduced by~\cite{chen2025predicting, comfortdataset} focuses on predicting driving comfort from environmental data, while \cite{meteier2023dataset} provides a multi-modal dataset that includes trust as one of several factors. The absence of publicly available datasets requires the initiation of new data collection campaigns for nearly every study, which are often costly, time-consuming, and logistically complex. Additionally, it impedes reproducibility and benchmarking, as researchers cannot directly compare their models or findings against prior work. Ultimately, this slows the development of effective models for understanding driver states in AVs.

\paragraph{ML-based trust and comfort estimation}
Many recent studies have utilized ML models to estimate driver trust or comfort in AVs. Researchers have explored diverse modalities for trust estimation, such as physiological data with Gradient Boosting, CNNs, or LSTMs models~\cite{ayoub2023real, yi2023measurement}, behavioral data with Kalman filters~\cite{azevedo2021real} and attention-based~\cite{zhu2023passenger}, hybrid~\cite{su2023estimating}, or reinforcement learning methods~\cite{xiang2022comfort} for comfort prediction. However, the absence of standardized datasets hinders reproducibility and cross-study comparison.

%%%%%%%%%%%%%%%%%%%%%%%%%%%%%%%%%%%%%%%%%%%%%%%%%%%%%%%%%%%%%%%%%%%%%%%%%%% Dataset %%%%%%%%%%%%%%%%%%%%%%%%%%%%%%%%%%%%%%%%%%%%%%%%%%%%%%%%%%%%%%%%%%%%%%%%%%%%

\section{Data Collection}
\subsection{Apparatus}
\paragraph{Driving simulator}

\begin{figure}[ht]
        \centering
        \includegraphics[width=0.95\linewidth]{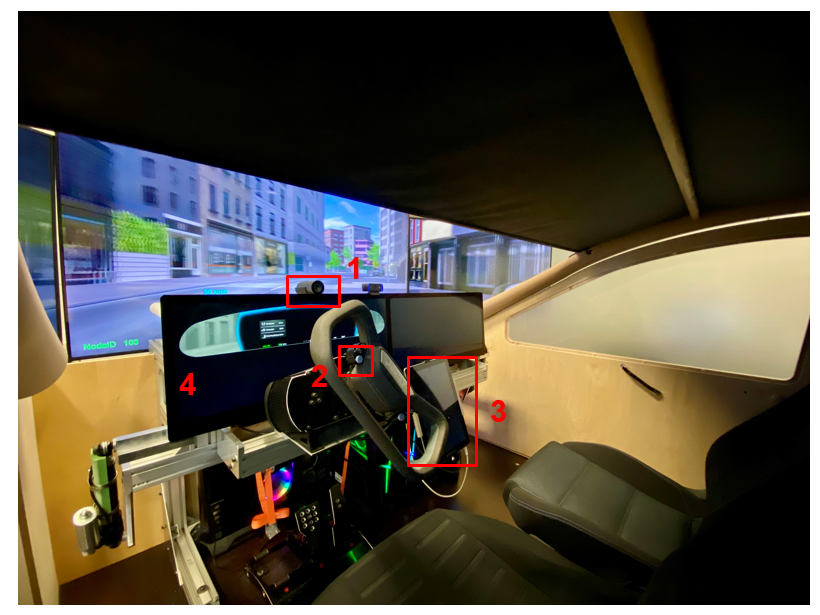}
        \vspace{-8pt}
        \caption{Stationary driving simulator with (1) gaze and emotion tracking camera, (2) button to turn on autonomous driving mode, (3) console to record trust voting, and (4) speedometer and side-view mirrors display}
        \label{fig:simulator}
   \end{figure}
   
The study was conducted in an immersive stationary driving simulator (see Fig.\@\ref{fig:simulator}) replicating a vehicle's interior. The setup included a driver and passenger seats, a steering wheel, and brake and accelerator pedals. During the manual driving test, participants used the steering wheel, which featured indicators and a button (see Fig.\@\ref{fig:simulator}(2)) for switching to fully autonomous driving. The side-view mirrors and speedometer were projected on the console above the steering wheel (see Fig.\@\ref{fig:simulator}(4)). The simulated environment was presented using three LCD screens and accompanying audio for an immersive experience. The automated driving scenario was developed using SILAB 7.2\footnote{https://wivw.de/en/silab-2/} driving simulation software. The environment variables, such as weather conditions, ego vehicle speed, and acceleration, positions, and velocities of surrounding vehicles, from the software were recorded at a sampling rate of $120Hz$.

\paragraph{Questionnaires}
Multiple pre- and post-drive questionnaires were utilized to capture participant's subjective assessments:
\begin{itemize}
    \item The Trust in Automation (TiA) questionnaire \cite{korber2018introduction} was administered using a 5-point Likert scale (1: "strongly disagree" and 5: "strongly agree") before and after the study to evaluate changes in trust.
    \item 11-point MIsery Scale (MISC) \cite{bos2005motion} was rated after both drives to gain insights into participants' subjective well-being and simulator sickness. The rating scale ranged from 0 ("no problems") to 10 ("vomiting"), with each incremental value representing increased physical discomfort, including sensations of discomfort, dizziness, and nausea.
    \item Karolinsk Sleepiness Scale (KSS) \cite{aakerstedt1990subjective} was administered on an 11-point scale (1: “extremely alert” to 10: “extremely sleepy, cannot stay awake”) after each drive to measure the degree of sleepiness. 
    \item At the end of the study, participants answered subjective questions regarding various aspects of the driver's emotions on a scale of 1-5, where  1 indicates "very low/ very unrealistic" and 5 indicates "very high/ very realistic." The questions addressed their feelings about the system's adaptation to their emotions, their acceptance of an interaction based on emotional processing, and the simulator's driving behavior, such as the degree of realism it offered and their willingness to drive AVs.
\end{itemize}

\paragraph{Physiological measurements}
\textit{Heart rate} was measured with a wearable Bluetooth device from Polar\footnote{\url{https://www.polar.com/de/products/accessories/polar-verity-sense}} at a sampling rate of $1Hz$. The device uses Photoplethysmography (PPG) to measure heartbeat in beats per minute (BPM). On the contrary, \textit{Gaze and emotion} data were extracted from real-time facial video using the FaceTrack software by Visage Technologies\footnote{\url{https://visagetechnologies.com/facetrack/}}, operating at $25Hz$. The input video was captured using an onboard camera shown in Fig.\@\ref{fig:simulator}(1).

\paragraph{Self-reported trust and comfort (Ground-truth)}
\begin{figure}[ht]
        \centering
        \includegraphics[width=0.9\linewidth]{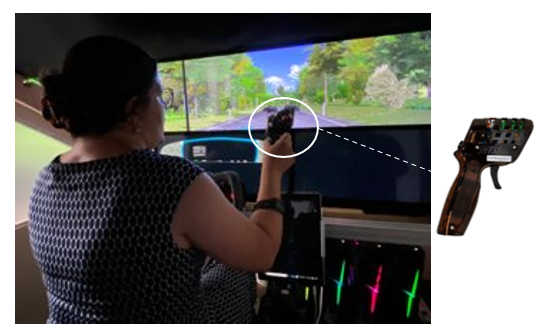}
        \vspace{-8pt}
        \caption{Participant holding the discomfort rating device during the drive}
        \label{fig:CRDI}
   \end{figure}
During each driving session, participants were asked to rate their level of trust after experiencing either a driving event or a neutral driving segment. They used a console (see Fig.\@\ref{fig:simulator}(3)) to provide ratings on a scale of 1-10, where 1 indicates 'low trust,' and 10 represents 'high trust.' Ten discrete trust ratings were recorded from each participant during the first drive, followed by seven trust ratings during the second drive. It is worth noting that, while a pop-up appeared on the console to record participants' current level of trust, the vehicle continued driving autonomously to reflect participants' immediate perceptions without interrupting the simulation flow.

In contrast, comfort data was continuously recorded using a handheld device\footnote{\url{https://wiki.frankenslot.de/anleitung/speedflow_triple}} (see Fig.\@\ref{fig:CRDI}) at a sampling rate of $300Hz$. Participants indicated their level of discomfort by pressing a lever on the device: a gentle press signified comfort, while a firm press indicated discomfort. The measurements were captured on a continuous scale ranging 0.0-100.0, with higher values representing greater discomfort.

\subsection{Experiment Design and Procedure}

Our study included two simulator-based fully autonomous driving sessions, each lasting approximately 11–12 minutes. Both drives began and ended at the exact location but followed different routes and events traversing urban and rural landscapes, mimicking real-world driving scenarios. Although the nature of events varied between the drives, both aimed to elicit dynamic trust and comfort responses across various conditions.

\begin{figure*}[ht]
    \centering
    \includegraphics[width=0.85\linewidth]{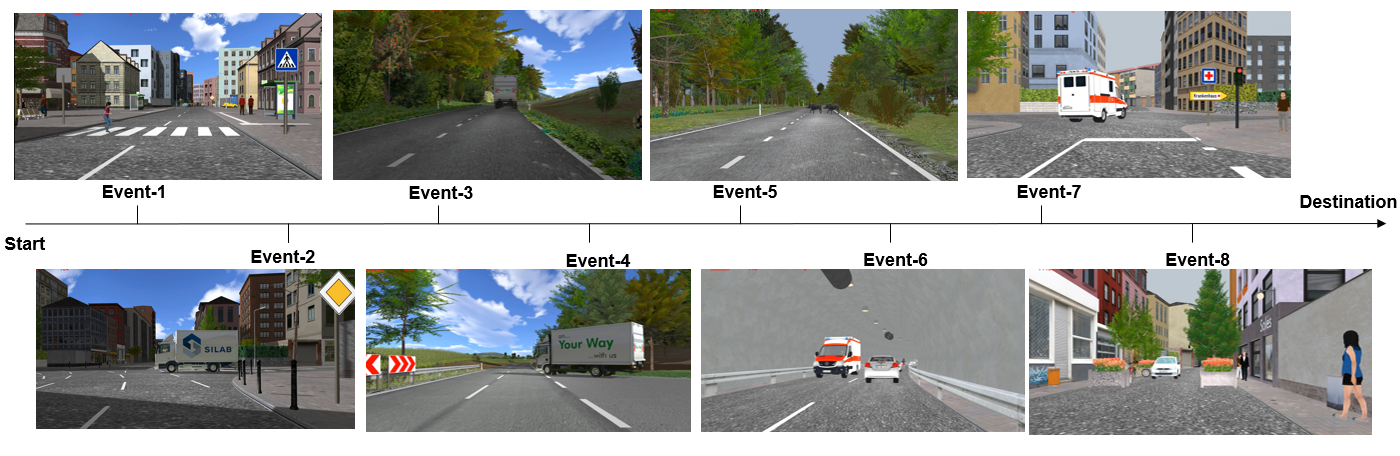}
    \vspace{-10pt}
    \caption{Driving events from first drive}
    \label{fig:drive1}
\end{figure*}

\begin{figure*}[ht]
\centering
    \includegraphics[width=0.85\linewidth]{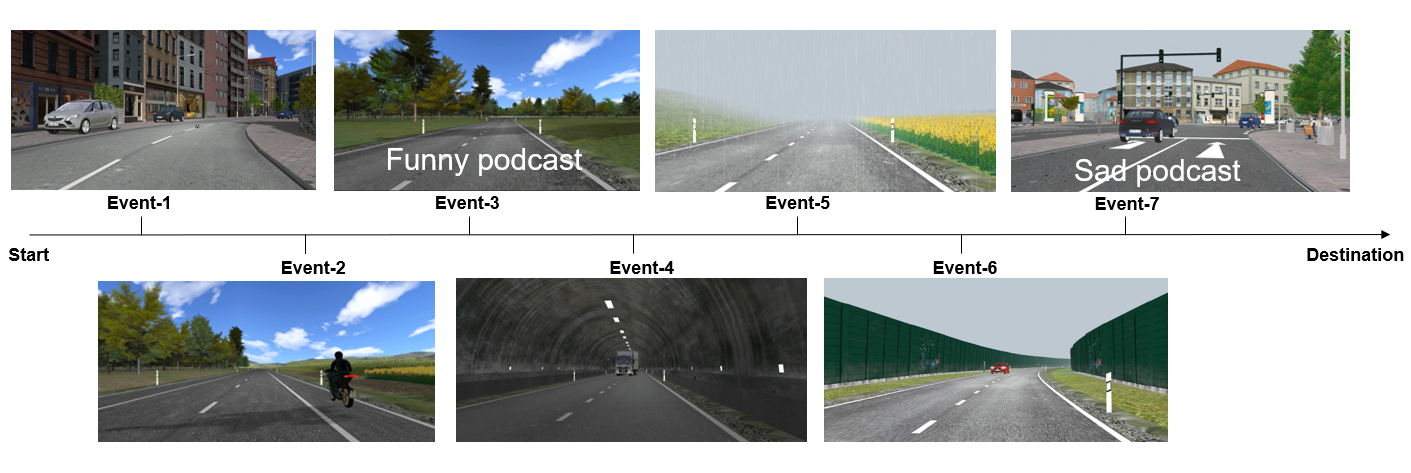}
    \vspace{-10pt}
    \caption{Driving events from second drive}
    \label{fig:drive2}
\end{figure*}

The first drive consisted of eight driving events, as shown in Fig.\@\ref{fig:drive1}, a pedestrian crossing the road, a truck at an intersection, a truck breakdown in the countryside, an unexpected truck overtaking, wildlife crossing the road, passing through a tunnel with an oncoming ambulance, the ambulance overtaking, and a narrow, obstructed road. In contrast, the second drive depicted in Fig.\@\ref{fig:drive2} featured seven events - a ball crossing the road, a motorbike overtaking the vehicle, a funny podcast playing on the radio, an oncoming truck in a tunnel, heavy rainfall accompanied by fog, a highway lined with noise barriers, and a sad podcast playing on the radio.

Upon arrival, participants were seated in the driving simulator and briefly introduced to the study. To help them familiarize with the controls and environment, each participant completed a short trial drive in manual mode. After the trial, participants initiated the experiment by pressing the automation button (see Fig.\@\ref{fig:simulator}(2)) to begin the first of two consecutive autonomous drives, without any manual driving in between. During each drive, we continuously recorded physiological data from sensors, comfort data reported by participants, and environmental data from the simulator. Additionally, we captured real-time discrete levels of trust reported by the participants. This design enabled us to temporally align the dynamic driving context and participants' physiological states with discrete trust assessments and continuous comfort responses for each drive and participant.

\subsection{Participants}
The study involved 31 participants (17 males and 14 females). This sample size was chosen to ensure sufficient statistical power for detecting within-subject effects, which refer to the variations in trust and comfort responses observed within the same individuals across different driving scenarios. 

Prior literature in human-automation or human-AV interaction~\cite{korber2018introduction} often reports medium to large within-subject effect sizes when examining factors such as trust, comfort, or physiological responses to driving events. These effect sizes (Cohen's d) typically range from 0.5 to 0.8. A power analysis conducted using medium effect size (Cohen's d $\approx$ 0.5), significance threshold ($\alpha$ = 0.05), and power ((1 - $\beta$) = 0.8) suggests that a minimum of 27–34 participants were required to achieve 80\% power in repeated-measures designs such as paired t-tests or within-subjects ANOVA. Therefore, the final sample size of 31 participants ensures that the study accounted for potential variability from continuous physiological measurements and multiple driving scenarios.

The participants' ages ranged from 18 to 66 years, with an average age of 41.4 years (\textit{SD} = 15.91, \textit{Med} = 35).  All participants, except for one, held a valid driving license.  Participants were monetarily compensated for their participation. No participants were excluded from the study.

%%%%%%%%%%%%%%%%%%%%%%%%%%%%%%%%%%%%%%%%%%%%%%%%%%%%%%%%%%%%%%%%%%%%%%%%%%% Dataset Processing %%%%%%%%%%%%%%%%%%%%%%%%%%%%%%%%%%%%%%%%%%%%%%%%%%%%%%%%%%%%%%%%%%%%%%%%%%%%

\section{TRUCE-AV Dataset}

\subsection{Data Cleaning and Synchronization}
At first, the raw data collected asynchronously from multiple sensors for physiological, environmental, and ground-truth measurements underwent structured data-cleaning. Each sensor stream was individually processed to handle missing values. Upon data cleaning, all sensor streams corresponding to a participant and a drive were synchronized using system-generated timestamps. The signals were resampled to align with the comfort rating data, which had the highest sampling frequency ($300 Hz$). The resampling of sensor data ensured that features were temporally aligned across various modalities. The cleaned and aligned sensor data were then consolidated into a combined file per participant per drive, enabling consistent access for downstream statistical modeling and machine learning tasks. Each of these combined files\footnote{The website includes a complete list of all the variables and their definitions} includes time-synchronized variables from all sensors. The following physiological features are used for trust and comfort estimation:
\begin{itemize}
    \item three gaze features representing the gaze-direction in x-, y-, and z-direction, respectively, with values ranging from -1.0 to 1.0. [variables: \textit{'visageData.VisageData.faceData.gazeDirection.i'}, where \mbox{$i\in\{0,1,2\}$}].
    \item seven emotion features representing the probabilistic emotional states such as anger, disgust, fear, happiness, sadness, surprise, and neutral, respectively, with values ranging from 0.0 to 1.0. [variables: \textit{'visageData.VisageData.analysisData.emotions.i'}, where \mbox{$i \in \{0, 1, 2, 3, 4, 5, 6\}$}]
    \item one heart rate feature recorded in BPM [variable: \textit{'polarData.PolarData.sensorData.heartRate'}]
\end{itemize}

The comfort ground truth ranging from 0.0 to 100.0 is recorded as \textit{'comfortGroundtruth.ComfortGroundtruth.measured\_comfort'} and discrete trust ground truth ranging from -1 to 10, where -1 signifies no trust vote as \textit{'silabData.SilabData.trustVote'}.
\subsection{Data statistics}

\paragraph{Questionnaires}

\begin{table}[ht]
\caption{Analysis of pre- and post-study TiA questionnaire}
\centering
    \begin{tabular}{p{2.5cm}p{1cm}p{0.7cm}p{1.2cm} p{0.5cm}}
    \toprule
         \textbf{Scale} & \textbf{pre-/ post-} & \textbf{Mean} & \textbf{Standard Deviation}  & \textbf{N}\\ 
    \midrule
         \multirow{2}{2.5cm}{\raggedright Intention of Developers}& pre-& 4.3& 0.76  &25\\
         & post- & 4.55&  0.5&   28\\ 
         \hline
         \multirow{2}{2.5cm}{\raggedright Propensity to Trust}& pre- & 3.24& 0.66 &28\\ 
         & post - & 3.37&  0.9& 28\\ 
         \hline
         \multirow{2}{2.5cm}{\raggedright Familiarity}&  pre- & 2.46&  1.37 &28\\ 
         & post- & 2.14&  1.25& 28\\ 
         \hline
        \multirow{2}{2.5cm}{\raggedright Reliability/ Competence}& pre- & - & - & -\\
        &post- & 3.51&  0.72& 25\\ 
        \hline
        \multirow{2}{2.5cm}{\raggedright Understanding/ Predictability}&  pre- & - & - & -\\
         &post &3.57&  0.88& 27\\ 
         \hline
        \multirow{2}{2.5cm}{\raggedright Trust in Automation}& pre- & - & - & -\\
         &post- & 3.73&  1.02& 30\\ 
    \bottomrule
    \end{tabular}   
     \label{tab:pre-collection TiA}
\end{table}

\begin{table}[ht]
    \centering
    \caption{Analysis of open questions collected at the end of study}    \label{tab:subjective question}
    \begin{tabular}{p{3.2cm}ccc}
    \toprule
         \textbf{Item} &  \textbf{Mean} &  \textbf{Standard Deviation} & \textbf{N} \\
    \midrule
         \multicolumn{1}{p{3.2cm}}{\raggedright Realism of Simulator}&  3.77&  0.76& 31\\ 
         \multicolumn{1}{p{3.2cm}}{\raggedright Intention to use autonomous driving}&  3.76&  1.12& 31\\ 
         \multicolumn{1}{p{3.2cm}}{\raggedright Acceptance of emotion processing} &  3.24&  1.42& 31\\ 
         \multicolumn{1}{p{3.2cm}}{\raggedright Acceptance of emotion adaption}&  3.15&  1.13& 31\\ 
    \bottomrule
    \end{tabular}
\end{table}

Table \@\ref{tab:pre-collection TiA} presents the analysis of the TiA questionnaire collected before and after the study. Further statistical analysis was conducted in RStudio to examine changes across the sub-scales:
\begin{itemize}
    \item \textit{Familiarity} significantly decreased, \mbox{$t(25) = -2.476$}, \mbox{$p = 0.020$}, \mbox{$d = 0.224$}. This decline may reflect participants' subjective interpretation of the term "similar system" in the item "I have experience with similar automated systems."
    \item  \textit{Intention of Developers} showed no significant change, \mbox{$t(26) = -1.344$}, \mbox{$p = 0.191$}, \mbox{$d = -0.321$}, suggesting that trust in the system's developers remained stable displaying consistent agreement with the item "I believe the developers of this system had the users' best interests in mind."
    \item \textit{Propensity to Trust} also remained unchanged, \mbox{$t(23) = -0.344$}, \mbox{$p = 0.734$}, \mbox{$d = -0.068$}, indicating that the test drive experience did not influence participant's general tendency to trust automation.
\end{itemize}

Here, $t(n)$ denotes the t-test with \textit{n} degrees of freedom, \textit{p} represents the p-value, and \textit{d} is Cohen's d. Mean ratings across the remaining sub-scales, Reliability/Competence, Understandability/Predictability, and Overall Trust (TiA), suggest mild agreement with statements such as "The automated system performs its tasks consistently and accurately," "I can anticipate how the system will behave in a given situation," and "I trust this automated system."  

The mean($\mu$) and standard deviation($\sigma$) of 11-point MISC~\cite{bos2005motion} from first and second drive are $\mu = 0.94$, $\sigma=1.36$ and $\mu = 1$, $\sigma=2.18$ respectively. Therefore, it can be concluded that most participants reported no problems or slight discomfort with no specific symptoms, with only a handful reporting mild dizziness or nausea. Further statistical analysis showed no difference in the subjective feeling of participants after both drives, with \mbox{$t(30) = -0.246$}, \mbox{$p = 0.0807$}, \mbox{$d = -0.031$}.

The $\mu$ and $\sigma$ of KSS~\cite{aakerstedt1990subjective}, for first and second drive are $\mu = 2.47$, $\sigma=1.35$ and $\mu = 2.43$, $\sigma=1.47$ respectively. Therefore, it can be concluded that most participants were alert, with only a few reporting signs of sleepiness. Further statistical analysis showed no difference in the sleepiness of participants after both drives, with \mbox{$t(15) = -0.460$}, \mbox{$p = 0.652$}, \mbox{$d = -0.087$}.

The mean ratings of open questions about subjective feelings towards the system and simulator are presented in Table \@\ref{tab:subjective question} suggesting a mild agreement to the sub-scales.

\paragraph{Trust and comfort ground truth}

\begin{figure}[ht]
    \centering
    \includegraphics[width=0.8\linewidth]{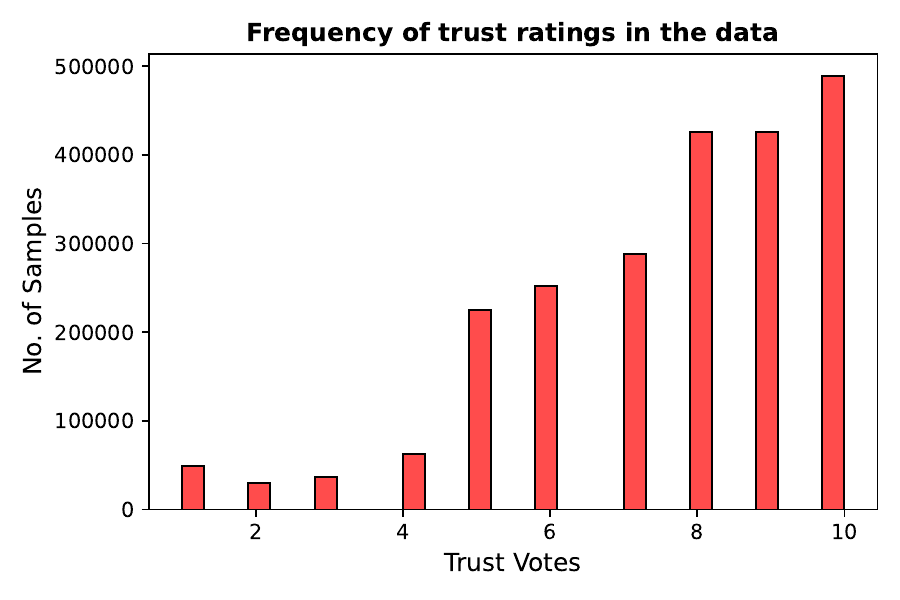}
    \vspace{-12pt}
    \caption{Distribution of collected trust ratings across all participants and drives}
    \label{fig:trust distribution}
\end{figure}

\begin{figure}[ht]
    \centering
    \includegraphics[width=0.8\linewidth]{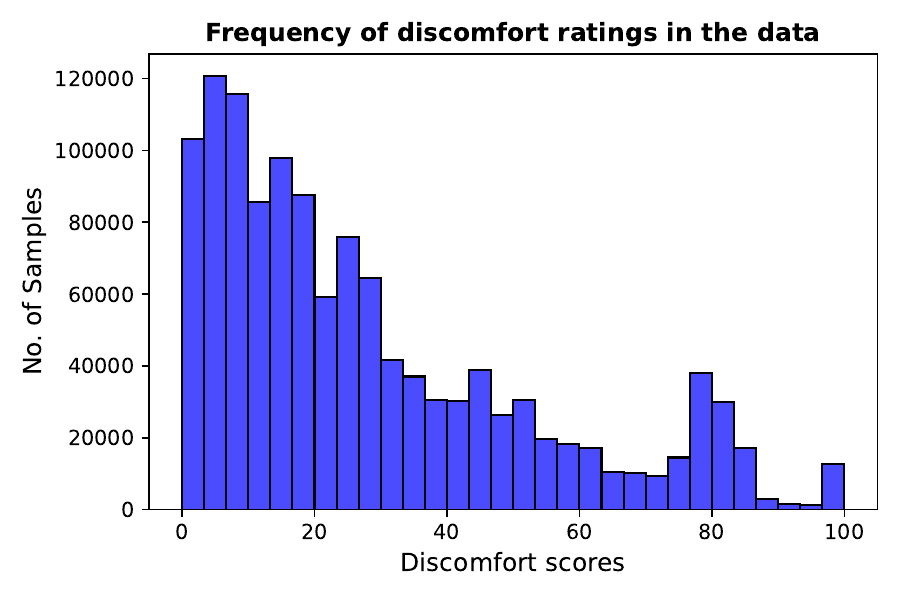}
    \vspace{-12pt}
    \caption{Distribution of recorded discomfort data across all participants and drives}
    \label{fig:comfort distribution}
\end{figure} 

Fig.\@\ref{fig:trust distribution} and Fig.\@\ref{fig:comfort distribution} present the distribution of trust and discomfort data across all participants and drives in the timeseries. Both distributions are skewed, primarily due to the frequent occurrence of high trust and comfort ratings among participants. The consistently high trust ratings observed during the driving sessions are consistent with participants' responses on the corresponding TiA subscale in the questionnaire. A strong positive correlation was identified between the TiA trust ratings and trust reported during drive-1 (\mbox{$\rho=0.725$}) and drive-2 (\mbox{$\rho=0.636$}), both of which showed high statistical significance (\mbox{$p<0.001$}).

Additionally, discomfort ratings displayed a strong negative correlation with trust during both driving sessions. Specifically, discomfort in drive-1 correlated negatively with trust in drive-1 (\mbox{$\rho=-0.733$}) and discomfort in drive-2 displayed an even stronger negative correlation with trust ratings from drive-2 (\mbox{$\rho=-0.974$}), also with high statistical significance (\mbox{$p=0.001$}). These findings indicate that as participants' trust in the system increased, their reported discomfort decreased, which supports the trends shown in the frequency distribution graphs for trust and comfort.

%%%%%%%%%%%%%%%%%%%%%%%%%%%%%%%%%%%%%%%%%%%%%%%%%%%%%%%%%%%%%%%%%%%%%%%%%%% Baseline Experiments %%%%%%%%%%%%%%%%%%%%%%%%%%%%%%%%%%%%%%%%%%%%%%%%%%%%%%%%%%%%%%%%%%%%%%%%%%%%
\section{Experiments}

\subsection{Training and Target Feature Extraction}

\paragraph{Training features} First, we implemented a modality-specific pre-processing pipeline for physiological data-based trust and comfort estimation to reduce inter-individual variability and ensure feature consistency. Heart rate was transformed using mean-based scaling $x' = ({x - \mu})/{C}$, where $\mu$ denotes the session-specific mean, and C is a fixed constant. The resulting values were then clipped within a predefined range to suppress outliers. Emotion features were mean-centered ($x' = x - \mu$) to preserve the relative differences while removing session-level bias. The gaze data were retained in their raw form to preserve directional information.

After normalization, we performed time-aligned feature extraction using a sliding-window approach. Let \( X \in \mathbb{R}^{L \times d} \) represent pre-processed physiological data, with \( L \) time steps and \( d \) features. For each trust or comfort label, we defined a window of $W$ milliseconds (ms). Given an average inter-sample interval \( \Delta t \), the number of samples available in each window was computed as $
T = \left\lfloor \frac{W}{\Delta t} \right\rfloor$.
From these $T$ samples, $N$ equally spaced samples $s_i$, where $i = 0, 1, \dots, N - 1$, were selected to ensure uniform coverage across the window. The feature extraction window moves by a stride of \( \delta \) ms, to collect $N$ samples, \( X_j^{N} = [X_{\delta + s_0}, \dots, X_{\delta + s_{N-1}}] \), at every window position $j$. Each sampled feature array \( X_j^{N} \in \mathbb{R}^{N \times d} \) was flattened into a feature vector \( x^{(j)} \in \mathbb{R}^{N \cdot d} \). This process was repeated for K windows traversing the complete data, each producing a structured input vector \( x^{(j)} \), to form the final training data $X^{train} \in \mathbb{R}^{K \times N\cdot d}$

\paragraph{Target features}
Similar to how training features were handled, the temporally aligned trust and comfort targets underwent pre-processing followed by sampling using a sliding window method to generate target data. The discrete trust labels, which included some values denoted as \(-1\) (indicating no response), were linearly interpolated to create a continuous signal. To address boundary issues, both forward- and backward-filling methods were applied. In contrast, comfort ratings were continuously recorded and did not require interpolation. After pre-processing, the trust label \( y_{t}^{(j)} \) and the comfort target \( y_{c}^{(j)} \) were sampled at the final index of the sliding window corresponding to a feature vector \( x^{(j)} \). A fixed scaling factor (e.g., 100) was applied to normalize the comfort targets. The sampling process was repeated for K windows, resulting in target vectors \( Y_{t}, Y_{c} \in \mathbb{R}^K \) for trust and comfort data, respectively, thereby creating a fully supervised dataset for the estimation of trust and comfort.

\subsection{Baseline Models for Trust and Comfort Estimation}

To evaluate the effectiveness of physiological data for estimating trust and comfort, we implemented a set of baseline models tailored to the nature of each task: multi-class classification for trust and regression for comfort. 

\paragraph{Trust Estimation - Multi-class classification}
Trust prediction was framed as a 10-class classification problem, and evaluated using standard classifiers across three categories: linear, tree-based, and nonlinear models. Performance was assessed using mean accuracy, macro-averaged F1-score, precision, and recall.

\paragraph{Comfort Estimation - Regression}
Comfort estimation was formulated as a regression task, with normalized continuous target values in the range [0, 1]. Models were evaluated across three categories: linear, tree-based, and nonlinear regressors. Performance was assessed using three metrics: Mean Absolute Error (MAE), Root Mean Squared Error (RMSE), and the coefficient of determination ($R^2$).

\paragraph{Implementation Details}
To generate training and target data, we set $W=10000ms$,  $N=20$, and $\delta=100ms$. All the models were implemented using the \texttt{scikit-learn} library using default parameters. In addition, a 5-fold cross-validation strategy was employed to ensure generalization. 

\section{Results}

\subsection{Trust Estimation and Feature Analysis}

Several models were evaluated for multi-class trust estimation using physiological data, the results are summarized in Table \@\ref{tab:model_performance}. Random Forest achieved the best performance across all metrics. Other tree-based models also performed competitively, with XGBoost reaching 82.64\% accuracy and LightGBM achieving 76.12\%  accuracy. In contrast, Linear models, such as Logistic Regression, Linear Support Vector Classifier (SVC) performed poorly overall. They showed particularly weak predictive performance on the metric F1-score, indicating that they struggle to address class imbalance and feature nonlinearity effectively.

\begin{table}[h]
\renewcommand{\arraystretch}{1.3}
    \centering
    \caption{Model performance comparison for trust estimation}
    \label{tab:model_performance}

    \begin{tabular}{p{1.2cm} p{1.3cm} p{1cm} p{0.7cm} p{0.8cm} p{0.8cm}}
        \hline 
        \textbf{Category} & \textbf{Model} & \textbf{Accuracy (Mean)} & \textbf{F1-score (Mean)} & \textbf{Precision (Mean)} & \textbf{Recall (Mean)}  \\
        \hline
        \multirow{4}{1.2cm}{Linear models} & \multicolumn{1}{p{1.3cm}}{\raggedright Logistic Regression} & 26.06\% & 10.24\% & 26.82\% & 12.67\%  \\
            & \multicolumn{1}{p{1.3cm}}{\raggedright LinearSVC} & 25.98\% & 9.29\% & 15.98\% & 12.34\% \\
            & \multicolumn{1}{p{1.3cm}}{\raggedright Ridge classifier} & 25.98\% & 9.27\% & 15.66\% & 12.34\% \\
            & \multicolumn{1}{p{1.3cm}}{\raggedright SGD classifier} & 20.58\% & 9.5\% & 12.22\% & 11.58\%  \\
            \hline
        \multirow{4}{1.2cm}{Tree-based models} & \multicolumn{1}{p{1.3cm}}{\raggedright Random Forest} & \textbf{94.42\%} & \textbf{93.73\%} & \textbf{96.18\%} & \textbf{91.61\%}  \\
            & \multicolumn{1}{p{1.3cm}}{\raggedright HistGradient Boosting} & 76.92\% & 76.34\% & 79.76\% & 73.6\% \\
            & \multicolumn{1}{p{1.3cm}}{\raggedright XGBoost} & 82.64\% & 83.49\% & 86.72\% & 80.83\%  \\
            & \multicolumn{1}{p{1.3cm}}{\raggedright LightGBM} & 76.12\% & 76.46\% & 79.63\% & 73.91\% \\
            \hline
        \multirow{2}{1.2cm}{Nonlinear/ other models} & \multicolumn{1}{p{1.3cm}}{\raggedright KNN} & 78.83\% & 74.27\% & 72.11\% & 77.49\% \\
            & \multicolumn{1}{p{1.3cm}}{\raggedright MLP classifier} & 51.68\% & 45.56\% & 46.48\% & 45.42\% \\
            \hline
    \end{tabular}
\end{table}

\begin{figure}[ht]
    \centering
    \includegraphics[width=0.8\linewidth]{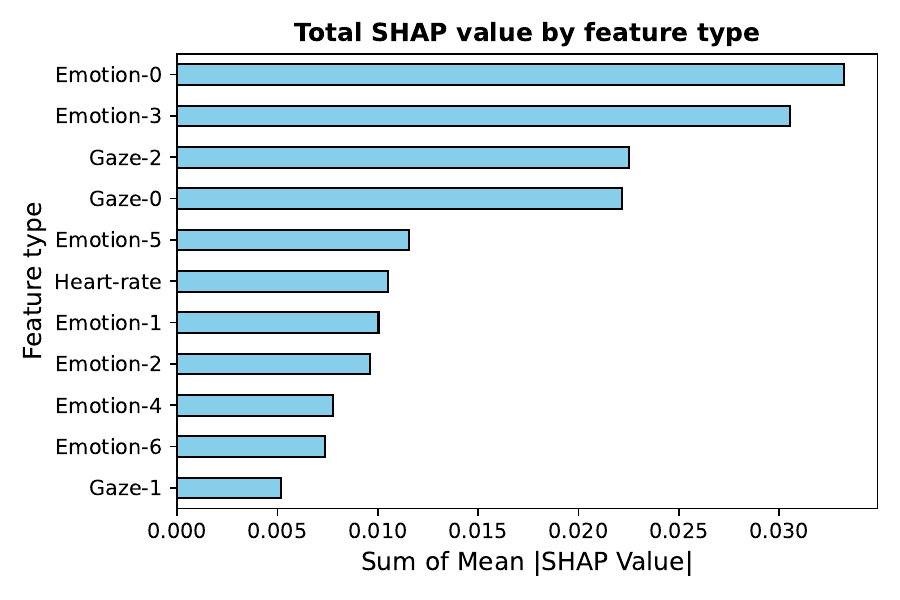}
    \vspace{-12pt}
    \caption{SHAP importance of individual features for trust estimation}
    \label{fig:shap analysis1}
\end{figure}

\begin{figure}[ht]
    \centering
    \includegraphics[width=0.8\linewidth]{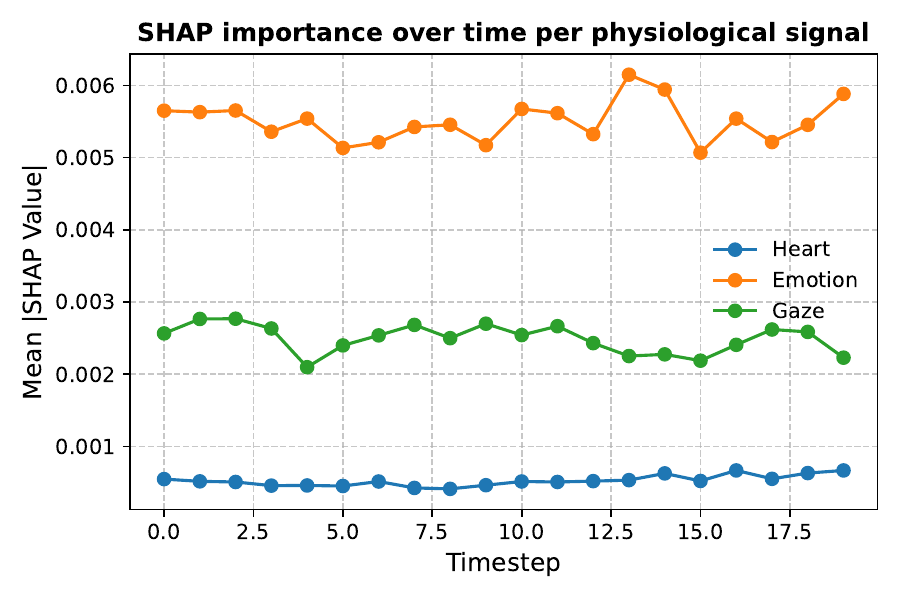}
    \vspace{-12pt}
    \caption{SHAP importance over time for each physiological signal for trust estimation}
    \label{fig:SHAP graph1}
\end{figure}

Additionally, we conducted feature importance analysis using SHAP on the Random Forest classifier, selected for its strong predictive performance. Fig.\@\ref{fig:shap analysis1} summarizes the impact of individual features on model outputs, and Fig.\@\ref{fig:SHAP graph1} summarizes the impact of each signal type over time on model outputs. Emotion features had the most significant influence throughout the entire time window, demonstrating a stable and consistently high impact on the model's predictions. Gaze had a moderate effect, whereas heart rate contributed the least, maintaining consistently low importance values. These findings suggest that emotion is the most reliable indicator for estimating trust, gaze offers additional insights, and heart rate provides only limited information.

\subsection{Comfort Estimation and Feature Analysis}
Table \@\ref{tab:regression_results_no_time} summarizes the regression results from various models for comfort estimation. Among all models, the MLP Regressor achieved the best predictive performance with an $R^2$ score of 0.1714. Tree-based models such as Random Forest and XGBoost also showed reasonable performance, with Random Forest achieving the lowest error metrics (MAE = 0.0404; RMSE = 0.0985), reflecting its effectiveness in capturing non-linear relationships. In contrast, linear models such as Ridge and Linear Regression underperformed, with $R^2$ values near zero, indicating limited capacity to model the underlying behavioral patterns in physiological data.

\begin{table}[ht]
\renewcommand{\arraystretch}{1.2}
    \centering
    \caption{Model performance comparison for comfort estimation}
    \label{tab:regression_results_no_time}

    \begin{tabular}{p{1.1cm} p{2.2cm} p{1cm} p{0.7cm} p{0.8cm}}
        \hline
        \textbf{Category} & \textbf{Model} & \textbf{R\textsuperscript{2} (Mean)} & \textbf{MAE (Mean)} & \textbf{RMSE (Mean)} \\
        \hline
        \multirow{2}{1.1cm}{Linear models} & Linear regression & 0.0106 & 0.0451 & 0.1072  \\
            & Ridge regression & 0.0107 & 0.0451 & 0.1072  \\
            \hline
        \multirow{3}{1.1cm}{Tree-based models} & Random Forest & 0.1633 & \textbf{0.0404}  & \textbf{0.0985}  \\
            & Gradient Boosting & 0.0781 & 0.0429  & 0.1034  \\
            & XGBoost & 0.1480 & 0.0414 & 0.0994  \\
            \hline
        \multirow{2}{1.1cm}{Nonlinear models}& SVR (RBF Kernel) & 0.1452 & 0.0716 & 0.0996  \\
            & MLP Regressor & \textbf{0.1714}  & 0.0536 & 0.0981 \\
        \hline
    \end{tabular}
\end{table}

\begin{figure}[ht]
    \centering
    \includegraphics[width=0.8\linewidth]{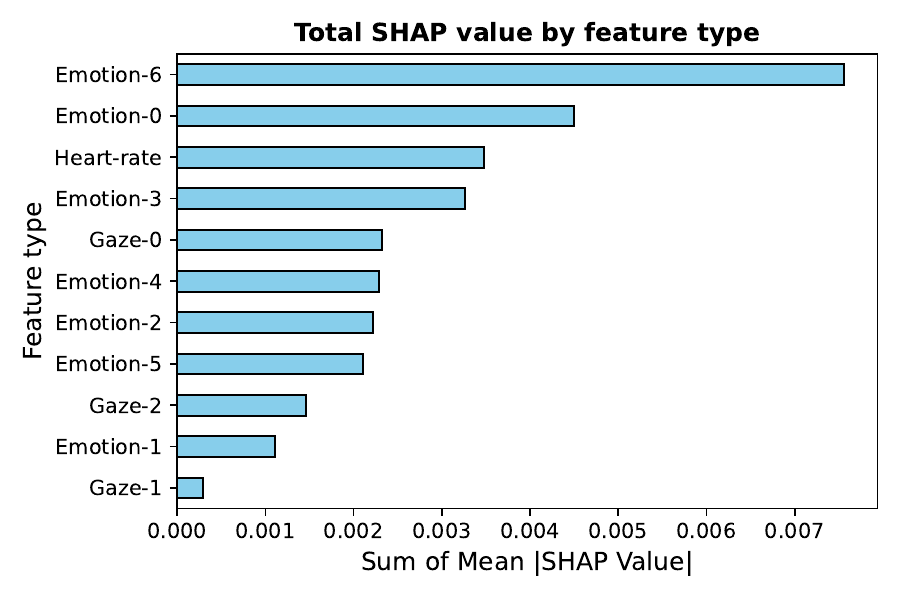}
    \vspace{-12pt}
    \caption{SHAP importance of individual features for comfort estimation}
    \label{fig:shap analysis}
\end{figure}

\begin{figure}[ht]
    \centering
    \includegraphics[width=0.8\linewidth]{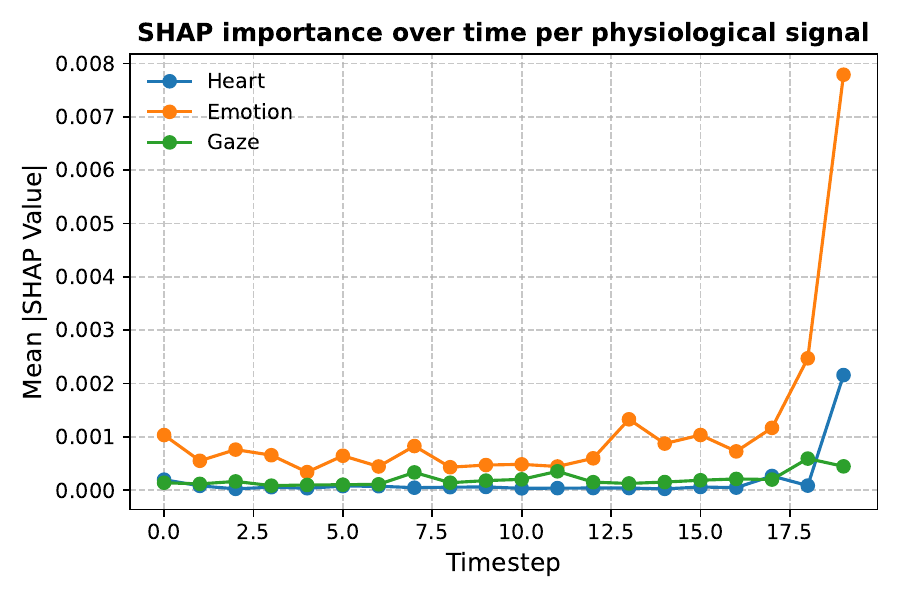}
    \vspace{-12pt}
    \caption{SHAP importance over time for each physiological signal for comfort estimation}
    \label{fig:SHAP graph}
\end{figure}

Additionally, we performed feature importance analysis using SHAP on the Random Forest Regressor, leveraging its ability to capture non-linear relationships in the data. Fig.\@\ref{fig:shap analysis} summarizes the impact of individual features on model outputs, and Fig.\@\ref{fig:SHAP graph} aggregates SHAP values over time for each physiological signal type. Emotion features consistently contribute more than other modalities, with a sharp rise in importance near the final time steps. This suggests that recent emotional cues are strong indicators of perceived comfort, while heart rate shows a moderate increase in influence toward the end. Gaze features, in contrast, exhibit consistently low SHAP values across time, implying limited predictive value in this setting.

%%%%%%%%%%%%%%%%%%%%%%%%%%%%%%%%%%%%%%%%%%%%%%%%%%%%%%%%%%%%%%%%%%%%%%%%

\section{Conclusion and Future Work}
This paper introduces the TRUCE-AV dataset, a novel multimodal dataset for estimating trust and comfort in AVs. Our dataset consists of continuous real-time modalities, such as physiological, environmental, trust votes, and comfort ratings, from 31 participants experiencing two immersive simulator-based fully autonomous driving sessions. To complement the real-time data, participants also completed questionnaires aimed at analyzing the relationship between dynamic (real-time) and static (survey-based) trust evaluations. Using standard baseline models, we evaluated the dataset's efficacy for both multi-class trust estimation and regression-based comfort prediction. Our comparative analysis revealed that tree-based and nonlinear models consistently outperformed linear approaches across both tasks, highlighting the complex, nonlinear nature of physiological responses associated with trust and comfort levels. Hence, the TRUCE-AV dataset is valuable for developing adaptive systems capable of dynamically responding to user trust and comfort levels, enhancing user acceptance and safety in AVs. 

While our work represents a significant step towards adaptive AV systems by introducing a publicly available dataset, several avenues for future research and improvement remain. Notably, our dataset's trust and comfort ratings distribution is skewed toward higher values, mirroring the conclusions drawn from questionnaires. This skewness may introduce bias and reduce model sensitivity to low-trust or high-discomfort states. Future work could address this imbalance by applying data augmentation or resampling techniques. Moreover, baseline models may not fully capture the dynamic and nonlinear nature of physiological responses, especially without temporal modeling. To better represent these dynamics, future research should investigate the use of RNNs, Temporal Convolutional Networks, or Transformer-based architectures to identify critical patterns over time. Lastly, while our current models rely solely on physiological signals for estimating trust and comfort, future research could incorporate environmental data from the dataset. By integrating environmental information, we can provide essential context for understanding physiological responses, leading to a more accurate and reliable estimations in dynamic, real-world conditions.

%%%%%%%%%%%%%%%%%%%%%%%%%%%%%%%%%%%%%%%%%%%%%%%%%%%%%%%%%%%%%%%%%%%%%%%%

%%% Use this environment to include acknowledgements (optional).
%%% This will be omitted in doubleblind mode.

\begin{ack}
This work is a result of the joint research project STADT:up (19A22006F). The project is supported by the German Federal Ministry for Economic Affairs and Energy (BMWE), based on a decision of the German Bundestag. The author is solely responsible for the content of this publication.
\end{ack}

%%%%%%%%%%%%%%%%%%%%%%%%%%%%%%%%%%%%%%%%%%%%%%%%%%%%%%%%%%%%%%%%%%%%%%%%

%%% Use this command to include your bibliography file.

\bibliography{mybibfile}

\begin{thebibliography}{41}
\providecommand{\natexlab}[1]{#1}
\providecommand{\url}[1]{\texttt{#1}}
\expandafter\ifx\csname urlstyle\endcsname\relax
  \providecommand{\doi}[1]{doi: #1}\else
  \providecommand{\doi}{doi: \begingroup \urlstyle{rm}\Url}\fi

\bibitem[{\AA}kerstedt and Gillberg(1990)]{aakerstedt1990subjective}
T.~{\AA}kerstedt and M.~Gillberg.
\newblock Subjective and objective sleepiness in the active individual.
\newblock \emph{International journal of neuroscience}, 52\penalty0 (1-2):\penalty0 29--37, 1990.

\bibitem[Aledhari et~al.(2023)Aledhari, Rahouti, Qadir, Qolomany, Guizani, and Al-Fuqaha]{aledhari2023motion}
M.~Aledhari, M.~Rahouti, J.~Qadir, B.~Qolomany, M.~Guizani, and A.~Al-Fuqaha.
\newblock Motion comfort optimization for autonomous vehicles: Concepts, methods, and techniques.
\newblock \emph{IEEE Internet of Things Journal}, 11\penalty0 (1):\penalty0 378--402, 2023.

\bibitem[Asarar(2021)]{asarar2021predicting}
K.~Asarar.
\newblock Predicting comfort in autonomous driving from vibration measurements using machine learning models, 2021.

\bibitem[Ayoub et~al.(2021{\natexlab{a}})Ayoub, Avetisyan, Makki, and Zhou]{ayoub2021investigation}
J.~Ayoub, L.~Avetisyan, M.~Makki, and F.~Zhou.
\newblock An investigation of drivers’ dynamic situational trust in conditionally automated driving.
\newblock \emph{IEEE Transactions on Human-Machine Systems}, 52\penalty0 (3):\penalty0 501--511, 2021{\natexlab{a}}.

\bibitem[Ayoub et~al.(2021{\natexlab{b}})Ayoub, Yang, and Zhou]{ayoub2021modeling}
J.~Ayoub, X.~J. Yang, and F.~Zhou.
\newblock Modeling dispositional and initial learned trust in automated vehicles with predictability and explainability.
\newblock \emph{Transportation research part F: traffic psychology and behaviour}, 77:\penalty0 102--116, 2021{\natexlab{b}}.

\bibitem[Ayoub et~al.(2023)Ayoub, Avetisian, Yang, and Zhou]{ayoub2023real}
J.~Ayoub, L.~Avetisian, X.~J. Yang, and F.~Zhou.
\newblock Real-time trust prediction in conditionally automated driving using physiological measures.
\newblock \emph{IEEE Transactions on Intelligent Transportation Systems}, 24\penalty0 (12):\penalty0 14642--14650, 2023.

\bibitem[Azevedo-Sa et~al.(2021)Azevedo-Sa, Jayaraman, Esterwood, Yang, Robert~Jr, and Tilbury]{azevedo2021real}
H.~Azevedo-Sa, S.~K. Jayaraman, C.~T. Esterwood, X.~J. Yang, L.~P. Robert~Jr, and D.~M. Tilbury.
\newblock Real-time estimation of drivers’ trust in automated driving systems.
\newblock \emph{International Journal of Social Robotics}, 13\penalty0 (8):\penalty0 1911--1927, 2021.

\bibitem[Beggiato et~al.(2019)Beggiato, Hartwich, and Krems]{beggiato2019physiological}
M.~Beggiato, F.~Hartwich, and J.~Krems.
\newblock Physiological correlates of discomfort in automated driving.
\newblock \emph{Transportation research part F: traffic psychology and behaviour}, 66:\penalty0 445--458, 2019.

\bibitem[Beggiato et~al.(2020)Beggiato, Rauh, and Krems]{beggiato2020facial}
M.~Beggiato, N.~Rauh, and J.~Krems.
\newblock Facial expressions as indicator for discomfort in automated driving.
\newblock In \emph{Intelligent Human Systems Integration 2020: Proceedings of the 3rd International Conference on Intelligent Human Systems Integration (IHSI 2020): Integrating People and Intelligent Systems, February 19-21, 2020, Modena, Italy}, pages 932--937. Springer, 2020.

\bibitem[Bellem et~al.(2018)Bellem, Thiel, Schrauf, and Krems]{bellem2018comfort}
H.~Bellem, B.~Thiel, M.~Schrauf, and J.~F. Krems.
\newblock Comfort in automated driving: An analysis of preferences for different automated driving styles and their dependence on personality traits.
\newblock \emph{Transportation research part F: traffic psychology and behaviour}, 55:\penalty0 90--100, 2018.

\bibitem[Boor et~al.(2024)Boor, Rizk, Gershon, Lee, Mehler, and Reimer]{JDPower2024study}
L.~Boor, K.~Rizk, P.~Gershon, C.~Lee, B.~Mehler, and B.~Reimer.
\newblock Consumer acceptance of fully automated, self-driving vehicles, {J.D. Power 2024 U.S. Mobility Confidence Index (MCI)} study.
\newblock {White Paper}, JD Power and MIT Advanced Vehicle Technology Consortium, 2024.
\newblock URL \url{https://pavecampaign.org/wp-content/uploads/2025/01/JDP-2024-US-MCI-Whitepaper.pdf}.

\bibitem[Bos et~al.(2005)Bos, MacKinnon, and Patterson]{bos2005motion}
J.~E. Bos, S.~N. MacKinnon, and A.~Patterson.
\newblock Motion sickness symptoms in a ship motion simulator: effects of inside, outside, and no view.
\newblock \emph{Aviation, space, and environmental medicine}, 76\penalty0 (12):\penalty0 1111--1118, 2005.

\bibitem[Carsten and Martens(2019)]{carsten2019can}
O.~Carsten and M.~H. Martens.
\newblock How can humans understand their automated cars? hmi principles, problems and solutions.
\newblock \emph{Cognition, Technology \& Work}, 21\penalty0 (1):\penalty0 3--20, 2019.

\bibitem[Cegarra et~al.(2025)Cegarra, Unrein, Andre, Mouton, and Navarro]{cegarra2025driving}
J.~Cegarra, H.~Unrein, J.-M. Andre, O.~Mouton, and J.~Navarro.
\newblock Driving among autonomous vehicles: The effect of initial trust and driving style on driving behaviors.
\newblock \emph{Transportation Research Part F: Traffic Psychology and Behaviour}, 112:\penalty0 99--110, 2025.

\bibitem[Chen et~al.(2025)Chen, Liu, Ni, Hai, Huang, Xu, Ling, Shen, Yu, Wang, et~al.]{chen2025predicting}
Z.~Chen, Y.~Liu, W.~Ni, H.~Hai, C.~Huang, B.~Xu, Z.~Ling, Y.~Shen, W.~Yu, H.~Wang, et~al.
\newblock Predicting driving comfort in autonomous vehicles using road information and multi-head attention models.
\newblock \emph{Nature Communications}, 16\penalty0 (1):\penalty0 2709, 2025.

\bibitem[Choi and Ji(2015)]{choi2015investigating}
J.~K. Choi and Y.~G. Ji.
\newblock Investigating the importance of trust on adopting an autonomous vehicle.
\newblock \emph{International Journal of Human-Computer Interaction}, 31\penalty0 (10):\penalty0 692--702, 2015.

\bibitem[De~Looze et~al.(2003)De~Looze, Kuijt-Evers, and Van~Dieen]{de2003sitting}
M.~P. De~Looze, L.~F. Kuijt-Evers, and J.~Van~Dieen.
\newblock Sitting comfort and discomfort and the relationships with objective measures.
\newblock \emph{Ergonomics}, 46\penalty0 (10):\penalty0 985--997, 2003.

\bibitem[Elbanhawi et~al.(2015)Elbanhawi, Simic, and Jazar]{elbanhawi2015passenger}
M.~Elbanhawi, M.~Simic, and R.~Jazar.
\newblock In the passenger seat: investigating ride comfort measures in autonomous cars.
\newblock \emph{IEEE Intelligent transportation systems magazine}, 7\penalty0 (3):\penalty0 4--17, 2015.

\bibitem[Hartwich et~al.(2021)Hartwich, Hollander, Johannmeyer, and Krems]{hartwich2021improving}
F.~Hartwich, C.~Hollander, D.~Johannmeyer, and J.~F. Krems.
\newblock Improving passenger experience and trust in automated vehicles through user-adaptive hmis:“the more the better” does not apply to everyone.
\newblock \emph{Frontiers in Human Dynamics}, 3:\penalty0 669030, 2021.

\bibitem[He et~al.(2022)He, Stapel, Wang, and Happee]{he2022modelling}
X.~He, J.~Stapel, M.~Wang, and R.~Happee.
\newblock Modelling perceived risk and trust in driving automation reacting to merging and braking vehicles.
\newblock \emph{Transportation research part F: traffic psychology and behaviour}, 86:\penalty0 178--195, 2022.

\bibitem[Hunter et~al.(2022)Hunter, Konishi, Jain, Akash, Wu, Misu, and Reid]{hunter2022interaction}
J.~G. Hunter, M.~Konishi, N.~Jain, K.~Akash, X.~Wu, T.~Misu, and T.~Reid.
\newblock The interaction gap: A step toward understanding trust in autonomous vehicles between encounters.
\newblock \emph{Proceedings of the Human Factors and Ergonomics Society Annual Meeting}, 66\penalty0 (1):\penalty0 147--151, 2022.

\bibitem[Kaufman(2025)]{kaufman2025improving}
R.~A. Kaufman.
\newblock \emph{Improving Human-Autonomous Vehicle Interaction in Complex Systems}.
\newblock PhD thesis, University of California, San Diego, 2025.

\bibitem[K{\"o}rber et~al.(2018)K{\"o}rber, Baseler, and Bengler]{korber2018introduction}
M.~K{\"o}rber, E.~Baseler, and K.~Bengler.
\newblock Introduction matters: Manipulating trust in automation and reliance in automated driving.
\newblock \emph{Applied ergonomics}, 66:\penalty0 18--31, 2018.

\bibitem[Lee and See(2004)]{lee2004trust}
J.~D. Lee and K.~A. See.
\newblock Trust in automation: Designing for appropriate reliance.
\newblock \emph{Human factors}, 46\penalty0 (1):\penalty0 50--80, 2004.

\bibitem[Meng et~al.(2024)Meng, Zhao, Chen, Wang, and Yu]{meng2024study}
H.~Meng, X.~Zhao, J.~Chen, B.~Wang, and Z.~Yu.
\newblock Study on physiological representation of passenger cognitive comfort: An example with overtaking scenarios.
\newblock \emph{Transportation research part F: traffic psychology and behaviour}, 102:\penalty0 241--259, 2024.

\bibitem[Meteier et~al.(2023)Meteier, Capallera, De~Salis, Angelini, Carrino, Widmer, Abou~Khaled, Mugellini, and Sonderegger]{meteier2023dataset}
Q.~Meteier, M.~Capallera, E.~De~Salis, L.~Angelini, S.~Carrino, M.~Widmer, O.~Abou~Khaled, E.~Mugellini, and A.~Sonderegger.
\newblock A dataset on the physiological state and behavior of drivers in conditionally automated driving.
\newblock \emph{Data in brief}, 47:\penalty0 109027, 2023.

\bibitem[Naiseh et~al.(2025)Naiseh, Clark, Akarsu, Hanoch, Brito, Wald, Webster, and Shukla]{naiseh2025trust}
M.~Naiseh, J.~Clark, T.~Akarsu, Y.~Hanoch, M.~Brito, M.~Wald, T.~Webster, and P.~Shukla.
\newblock Trust, risk perception, and intention to use autonomous vehicles: an interdisciplinary bibliometric review.
\newblock \emph{AI \& society}, 40\penalty0 (2):\penalty0 1091--1111, 2025.

\bibitem[Norzam et~al.(2022)Norzam, Karjanto, Yusof, Hassan, Zulkifli, Ab~Rashid, et~al.]{norzam2022analysis}
M.~N. A.~M. Norzam, J.~Karjanto, N.~M. Yusof, M.~Z. Hassan, A.~F.~H. Zulkifli, A.~A. Ab~Rashid, et~al.
\newblock Analysis of user’s comfort on automated vehicle riding simulation using subjective and objective measurements.
\newblock \emph{Automotive experiences}, 5\penalty0 (2):\penalty0 238--250, 2022.

\bibitem[Paschalidis et~al.(2020)Paschalidis, Hajiseyedjavadi, Wei, Solernou, Jamson, Merat, Romano, and Boer]{paschalidis2020deriving}
E.~Paschalidis, F.~Hajiseyedjavadi, C.~Wei, A.~Solernou, A.~H. Jamson, N.~Merat, R.~Romano, and E.~R. Boer.
\newblock Deriving metrics of driving comfort for autonomous vehicles: A dynamic latent variable model of speed choice.
\newblock \emph{Analytic methods in accident research}, 28:\penalty0 100133, 2020.

\bibitem[Peng et~al.(2024)Peng, Carlowitz, Madigan, Marberger, Lee, Krems, Beggiato, Romano, Wei, et~al.]{peng2023conceptualising}
C.~Peng, S.~Carlowitz, R.~Madigan, C.~Marberger, J.~D. Lee, J.~Krems, M.~Beggiato, R.~Romano, C.~Wei, et~al.
\newblock Conceptualising user comfort in automated driving: Findings from an expert group workshop.
\newblock \emph{Transportation Research Interdisciplinary Perspectives}, 24:\penalty0 101070, 2024.
\newblock \doi{https://doi.org/10.1016/j.trip.2024.101070}.

\bibitem[Siebert et~al.(2013)Siebert, Oehl, H{\"o}ger, and Pfister]{siebert2013discomfort}
F.~W. Siebert, M.~Oehl, R.~H{\"o}ger, and H.-R. Pfister.
\newblock Discomfort in automated driving--the disco-scale.
\newblock In \emph{HCI International 2013-Posters’ Extended Abstracts: International Conference, HCI International 2013, Las Vegas, NV, USA, July 21-26, 2013, Proceedings, Part II 15}, pages 337--341. Springer, 2013.

\bibitem[Stapel et~al.(2022)Stapel, Gentner, and Happee]{stapel2022road}
J.~Stapel, A.~Gentner, and R.~Happee.
\newblock On-road trust and perceived risk in level 2 automation.
\newblock \emph{Transportation research part F: traffic psychology and behaviour}, 89:\penalty0 355--370, 2022.

\bibitem[Su and Jia(2021)]{su2021study}
H.~Su and Y.~Jia.
\newblock Study of human comfort in autonomous vehicles using wearable sensors.
\newblock \emph{IEEE Transactions on Intelligent Transportation Systems}, 23\penalty0 (8):\penalty0 11490--11504, 2021.

\bibitem[Su and Jia(2023)]{su2023estimating}
H.~Su and Y.~Jia.
\newblock Estimating human comfort levels in autonomous vehicles based on vehicular behaviors and physiological signals.
\newblock In \emph{2023 IEEE/RSJ International Conference on Intelligent Robots and Systems (IROS)}, pages 9865--9870. IEEE, 2023.

\bibitem[Wang et~al.(2020)Wang, Zhang, Huang, and Zhao]{wang2020safety}
J.~Wang, L.~Zhang, Y.~Huang, and J.~Zhao.
\newblock Safety of autonomous vehicles.
\newblock \emph{Journal of advanced transportation}, 2020\penalty0 (1):\penalty0 8867757, 2020.

\bibitem[Xiang and Guo(2022)]{xiang2022comfort}
J.~Xiang and L.~Guo.
\newblock Comfort improvement for autonomous vehicles using reinforcement learning with in-situ human feedback.
\newblock Technical report, SAE Technical Paper, 2022.

\bibitem[Yi et~al.(2023{\natexlab{a}})Yi, Cao, Song, Wang, Guo, and Huang]{yi2023human}
B.~Yi, H.~Cao, X.~Song, J.~Wang, W.~Guo, and Z.~Huang.
\newblock How human-automation interaction experiences, trust propensity and dynamic trust affect drivers’ physiological responses in conditionally automated driving: Moderated moderated-mediation analyses.
\newblock \emph{Transportation research part F: traffic psychology and behaviour}, 94:\penalty0 133--150, 2023{\natexlab{a}}.

\bibitem[Yi et~al.(2023{\natexlab{b}})Yi, Cao, Song, Wang, Guo, and Huang]{yi2023measurement}
B.~Yi, H.~Cao, X.~Song, J.~Wang, W.~Guo, and Z.~Huang.
\newblock Measurement and real-time recognition of driver trust in conditionally automated vehicles: Using multimodal feature fusions network.
\newblock \emph{Transportation research record}, 2677\penalty0 (8):\penalty0 311--330, 2023{\natexlab{b}}.

\bibitem[Zhang et~al.(2022)Zhang, Tian, and Duffy]{zhang2022trust}
Z.~Zhang, R.~Tian, and V.~G. Duffy.
\newblock Trust in automated vehicle: A meta-analysis.
\newblock In \emph{Human-automation interaction: Transportation}, pages 221--234. Springer, 2022.

\bibitem[Zheng and Shyrokau(2023)]{comfortdataset}
Y.~Zheng and B.~B. Shyrokau.
\newblock Comfort-oriented driving: performance comparison between human drivers and motion planners (dataset), 2023.
\newblock URL \url{https://data.4tu.nl/articles/_/21947618/1}.

\bibitem[Zhu et~al.(2023)Zhu, Xi, Hu, Zhao, and Niu]{zhu2023passenger}
W.~Zhu, Z.~Xi, C.~Hu, B.~Zhao, and Y.~Niu.
\newblock Passenger comfort quantification for automated vehicle based on stacking of psychophysics mechanism and encoder-transformer model.
\newblock \emph{IEEE Transactions on Intelligent Transportation Systems}, 25\penalty0 (6):\penalty0 5211--5224, 2023.

\end{thebibliography}

\end{document}